\documentclass[aps,preprint,floatfix,nofootinbib,showpacs]{revtex4-1}
\pdfoutput=1
\usepackage{graphicx,color,float, amsmath}
\usepackage{hyperref}
\usepackage{MnSymbol}%
\usepackage{wasysym}%
\usepackage{color}

\begin{document}

{\small
\begin{flushright}
\end{flushright} }

\title{Lepton Universality Violation by Kaluza-Klein Neutrinos in $b\to s l l$ transition}
\vspace*{1cm}

\author{
Janus~Capellan~Aban, Chuan-Ren Chen, and Chrisna~Setyo~Nugroho}
\affiliation{
\vspace*{.5cm}
Department of Physics, National Taiwan Normal University, Taipei 116, Taiwan 
\vspace*{1cm}}
\date{\today}

\begin{abstract}
Recent measure of $R_K$, involving the decays of $b\to s l^+ l^-$, by the LHCb at CERN strengthened the deviation from the Standard Model prediction. The best fit in the updated global analysis suggests that the muon specific Wilson coefficients $C^{bs\mu \mu}_9=-C^{bs\mu \mu}_{10}$ should be about $-0.41$. In this paper, we show that the accumulate effects of KK modes of a singlet Dirac neutrino propagating in the large extra-dimensional space naturally provide $C^{bs l l}_9=-C^{bs l l}_{10}$ to explain the anomaly.  By taking the muon Yukawa coupling to be of ${\cal O}(1)$, the fundamental scale in the extra-dimensional framework should be lowered down to about $2.9$ TeV if there are two additional spatial dimensions. \
\end{abstract}

\maketitle

\section{Introduciton}

Even though the Standard Model (SM) of particle physics well explains most of the experimental results, it is widely believed that SM is an incomplete  theory. Theoretically, questions that remain unanswered by the SM stimulate the development of new theory beyond the SM, e.g. naturalness/hierarchy problem spurs supersymmetric and extra-dimensional frameworks. Observations that deviate from SM predictions, such as nonzero masses for neutrinos and  astronomical evidence for dark matter, usually call for extension of particle spectrum. 
Discoveries of any signals beyond the SM would be crucial to figure out what the new theory should be.   

Recently, the measurements of B hadron rare decay branching ratios $Br(B^+ \to K^+ e^+ e^- )$ and  $Br(B^+\to K^+ \mu^+ \mu^-)$ were updated by LHCb collaboration and the ratio, defined as $R_K\equiv Br(B^+ \to K^+ e^+ e^- )/Br(B^+\to K^+ \mu^+ \mu^-)=0.846$, is $3.1\sigma$~\cite{LHCb:2021trn} away from the SM expectation $R_K=1$~\cite{Greub:2008cy}, showing the hint of violation for lepton universality in the SM.  
Theoretically, the decays mentioned above involve process $b\to s l^+ l^-$ in quark level, and it can be parametrized in the well-established effective Hamiltonian  approach. The relevant operators are ${\cal O}_9$ and ${\cal O}_{10}$ with the associated Wilson coefficients $C_9$ and $C_{10}$, respectively. 
Actually, the rare B decays with anomalies have been studied for decades, including $B^0\to K^{*0}  l^+ l^-$, $B^\pm \to K^{*^\pm}  l^+ l^-$, $B_s\to \mu^+\mu^-$, $B_s\to \phi \mu^+\mu^-$. 
The updated global fit~\cite{Altmannshofer:2021qrr} study shows that, the best fit with $C_9^{bs\mu\mu}=-C_{10}^{bs\mu\mu}\simeq -0.39$ reaches the largest pull about $5.6 \sigma$ from the SM\footnote{After taking into account the data from Ref.~\cite{LHCb:2021lvy}, the largest pull is updated to be $5.9\sigma$ for the best fits $C_9^{bs\mu\mu}=-C_{10}^{bs\mu\mu}\simeq -0.41$~\cite{talk}}. 
Such a big discrepancy attracts much attention and new physics models have been proposed
~\cite{Fajfer:2012vx},
including the leptoquark~\cite{Hiller:2021pul}, a non-SM scalar that interacts with lepton and quark.  

In this paper, we revisit an extra-dimensional model in which a singlet Dirac neutrino is introduced and propagates in the bulk~\cite{Ioannisian:1999cw}. As a result, the active neutrinos receive tiny masses and mix with KK neutrinos after diagonalizing the mass matrix. The KK neutrinos then contribute to $C_9$ and $C_{10}$ through box diagrams, as to be discussed later. We further assume that significant contributions from KK neutrinos occur in the $\mu^\pm$ final state while the effect to electron mode is highly suppressed for simplicity.

The rest of this paper is organized as follows. In section~\ref{sec:model}, we briefly reviewed the extra-dimensional model discussed in~\cite{Ioannisian:1999cw}. 
The main analysis will be shown in section~\ref{sec:numerical}. We calculate the Wilson coefficients $C_9^{bs\mu\mu}=-C_{10}^{bs\mu\mu}$ after  suming over the contributions from KK neutrinos. We also consider the relevant constraints from lepton flavor physics, mainly focusing on $\mu \to e \gamma$ decay.  The parameter space that explains $C_9^{bs\mu\mu}=-C_{10}^{bs\mu\mu}=-0.41$ preferred by the global fit will be presented. Finally, we give our conclusions in section~\ref{sec:con}.

 \section{Model}
 \label{sec:model}

The tiny masses for neutrinos can be naturally generated in extra-dimensional framework~\cite{Hamed,Dienes,Dvali}. 
We revisit   
a model that minimally extends the SM by introducing a singlet Dirac neutrino $N(x,y)$. The field $N(x,y)$ is assumed to propagate in the compact extra dimensions in addition to the $3+1$ flat Minkowski spacetime, or in other words, $N(x,y)$ lives in a $(4+\delta)$-dimensional space, where $\delta$ counts for number of extra dimension~\cite{Ioannisian:1999cw}. As a convention, we denote the four-vector $x\equiv x^{\mu}$ where $\mu=0,1,2,3$, such that $x^0$ is the time coordinate and $x^i$ are the spatial coordinates for $i=1,2,3$. We also define the coordinates $y\equiv y^k$ with $k=1,2,..., \delta$ to describe the new extra dimensions. Under the periodic identification, $y \equiv y+2\pi R$, we compactify the coordinates $y$ on a circle with radius $R$. The Planck scale $M_P\simeq 10^{19}$~GeV and the fundamental scale $M_F$ can be related as
$M_F^{\delta+2}\simeq M_P^2 R^{-\delta}$. 
In the large extra dimension scenario, $\delta$ is required to be $\delta \ge 2$~\cite{Lella}. In the following, we briefly review the model given in~\cite{Ioannisian:1999cw}. 

For a simple illustration, we consider only the presence of one new large extra dimension $y$ in the Dirac field $N(x,y)$, since this can be naturally extended by applying the same compactification method to the case of higher extra dimensions. In lepton sector, SM  doublets and isosinglet Dirac neutrinos can be written as
\begin{align}
\label{eq:Doublet}
L_l (x)=
\begin{pmatrix}
\nu_{l}(x) \\
l (x)
\end{pmatrix}_{L}
\qquad \textrm{and} \qquad
N(x,y)=
\begin{pmatrix}
\xi(x,y) \\
\bar{\eta}(x,y)
\end{pmatrix}\, ,
\end{align}
where $\xi$ and $\bar{\eta}$ are both 5-dimensional two-component spinors, and $l=e,\mu,\tau$. 
The representations used for 5-dimensional gamma matrices are
\begin{align}
\label{eq:gmatrix}
\gamma_\mu =\begin{pmatrix}
0 & \bar{\sigma}_\mu\\
\sigma_\mu & 0
\end{pmatrix}
\qquad \textrm{and} \qquad 
\gamma_4 =
i\begin{pmatrix}
1_{2\times 2}& 0\\
0 & -1_{2\times2}
\end{pmatrix}
\end{align}
 where $\sigma^\mu=(1, \vec{\sigma})$ and $\bar{\sigma}^\mu=(1, -\vec{\sigma})$ such that $\vec{\sigma}=(\sigma_1,  \sigma_2, \sigma_3)$ with $\sigma_i$ being the Pauli matrices. The effective Lagrangian is given by
\begin{equation}
\label{eq:effLag}
\begin{aligned}
\mathcal{L}_{eff} = &\int_{0}^{2 \pi R} dy [\bar{N} (i \gamma^\mu \partial_\mu + i\gamma_4\partial_y)N - m\bar{N}N + \delta (y-a) \big( \sum_{e, \mu, \tau} 
\bar{h}_l L_l \tilde{\Phi}\xi + h.c.     \big) \\
&+ \delta(y-a)\mathcal{L}_{SM} (\Phi) ]\, , 
\end{aligned}
\end{equation}
where $y=a$ is the location of the intersection between $N(x,y)$ and 4D spacetime~\cite{APilaftsis:1999}, $\tilde{\Phi}=i\sigma_2 \Phi^*$ with $\Phi$ being the Higgs doublet, and $\mathcal{L}_{ SM } (\Phi)$ is the Lagrangian of the SM which is localized on 3-dimensional spatial space. The Yukawa coupling $\bar{h}_l$ is dimensionful and can be related to the dimensionless Yukawa coupling $h_l$,  as $\bar{h}_l =\dfrac{h_l}{{(M_F)}^{\delta/2}}$ for each $l=e,\mu, \tau$ particularly with $\delta=1$ in Eq.~(\ref{eq:effLag}).  
Now suppose that the field $N(x,y)$ in coordinate $y$ is $2\pi R$-periodic, i.e. $N(x,y)=N(x, y+ 2\pi R)$, then we can express its spinor components $\xi$ and $\eta$ in Fourier modes as:
\begin{align}
\label{eq:1stCom}
\xi(x,y)=\dfrac{1}{\sqrt{2\pi R}}\sum_{n=-\infty}^{\infty} \xi_{n}(x)\exp\big(\frac{iny}{R} \big) 
\end{align}
\begin{align}
\label{eq:2ndCom}
\eta(x,y)=\dfrac{1}{\sqrt{2\pi R}}\sum_{n=-\infty}^{\infty} \eta_{n}(x)\exp\big(\frac{iny}{R} \big)\,.
\end{align}

Substituting the Fourier expansion of $N(x,y)$ in the effective Lagrangian ~\eqref{eq:effLag} and further integrating over $y$, we obtain
\begin{align}
\label{eq:effLagRes}
\mathcal{L}_{eff} = &\mathcal{L}_{SM} + \sum_{-\infty}^{\infty} \Big\{ 
\bar{\xi}_n \big(i\bar{\sigma}^\mu \partial_\mu\big)\xi_n + \bar{\eta}_n \big(i\bar{\sigma}^\mu \partial_\mu\big)\eta_n - \Big[\Big(m+\frac{in}{R}\Big)\xi_n \eta_{-n}+ h.c.   \Big]  \nonumber      \\ 
& +\Big( \sum_{l=e, \mu, \tau} \bar{h}_{l}^{(n)}L_l
\tilde{\Phi}\xi_n + h.c.    \Big)     \Big\}\,,
\end{align}
where $\bar{h}_{l}^{(n)}=\dfrac{M_F}{M_P} h_l \exp \Big( \dfrac{ina}{R}\Big)$ 
for any $l=e,\mu, \tau$. The 4D Yukawa coupling $\bar{h}_{l}^{(n)}$  contains an automatic suppression by volume factor $\dfrac{M_F}{M_P}$ of the extra compactified dimensions, as in \cite{Hamed, Dienes}.\newline

Finally, the Lagrangian of KK neutrino mass matrix can be obtained from $\mathcal{L}_{eff}$ as
\begin{align}
\label{eq:massLag}
\mathcal{L}_{mass}^{KK}=\Psi_{+}^{T} \mathcal{M}\Psi_- + \text{h.c.}  \,.
\end{align}
After spontaneous symmetry breaking (SSB) of the Higgs doublet $\Phi$, in the basis of 
\begin{align}
\label{eq:weakbasis}
\Psi_{+}^T=(\nu_{iL}, \eta_0, \eta_{1},\eta_{-1},...,\eta_{n},\eta_{-n},... ) \quad
\textrm{and}\quad
 \Psi_{-}^T=(\xi_0, \xi_{-1},\xi_{1},..., \xi_{-n},\xi_{n},... )\,, 
\end{align}
the mass matrix reads
\begin{align}
\label{eq:massMatrix}
\mathcal{M} =\begin{pmatrix}
m_{l}^{(0)} & m_{l}^{(-1)}& m_{l}^{(1)}&\cdots&m_{l}^{(-n)}&m_{l}^{(n)}& \cdots\\
m& 0& 0& \cdots& 0& 0 & \cdots\\
0& m-\frac{i}{R}& 0& \cdots& 0& 0&... \\
0& 0& m+\frac{i}{R}& \cdots& 0& 0& \cdots \\
\vdots& \vdots& \vdots& \ddots& \vdots& \vdots& \cdots \\
0& 0& 0& \cdots& m-\frac{in}{R}& 0& \cdots \\
0& 0& 0& \cdots& 0& m+\frac{in}{R}& \cdots \\
\vdots& \vdots& \vdots& \vdots&& \vdots&  \ddots
\end{pmatrix}\,,
\end{align}
where $m_{l}^{(n)}=\bar{h}_{l}^{(n)}\, \frac{v}{\sqrt{2}}$ for each $l=e, \mu, \tau$ and
$\left\langle \Phi \right \rangle = \frac{v}{\sqrt{2}}$ with $v=246$ GeV. It is important to note that there are three more rows from the rectangular part of the neutrino mass matrix $\mathcal{M}$. These rows correspond to the massless Weyl spinors denoted by $\nu_{l(L)}^{T}=(\nu_{e(L)}, \nu_{\mu (L)}, \nu_{\tau(L)})$ and can be treated independently in the neutrino mass matrix $\mathcal{M}$. Moreover, $\nu_l$ are predominantly left-handed and thus refer to observable neutrinos.

To diagonalze the mass matrix, 
as prescribed in \cite{Ioannisian:1999cw}, one starts
with a new basis $\Psi_{+}^{R}$ which generates the weak basis $\Psi_+$ by a rotation using the unitary transformation $U^\nu$, i.e. 
$\Psi_+=U^\nu \Psi_{+}^{R}$ where $(\Psi_{+}^{R})^T=(\nu_l,\eta_{0}^{R}, \eta_{1}^{R},\eta_{-1}^{R},...,\eta_{n}^{R},\eta_{-n}^{R},... )$ with the superscript $R$ denoting the rotated basis. 
This unitary matrix is given by~\cite{Ioannisian:1999cw} 
\begin{align}
\label{eq:uniMatrix}
U^\nu =\begin{pmatrix}
( 1_{3\times 3} + \Xi^* \Xi^T )^{-1/2} & \Xi^*( 1 + \Xi^T \Xi^*)^{-1/2} \\
-\Xi^T( 1_{3\times 3} + \Xi^* \Xi^T )^{-1/2} & ( 1 + \Xi^T \Xi^*)^{-1/2}
\end{pmatrix}\,,
\end{align}
where
\begin{align}
\label{eq:KKLep}
\Xi=\big( \dfrac{m_{l}^{(0)}}{m}, \dfrac{m_{l}^{(-1)}}{m-\frac{i}{R}}, \dfrac{m_{l}^{(1)}}{m+\frac{i}{R}},...,\dfrac{m_{l}^{(-n)}}{m-\frac{in}{R}} ,\dfrac{m_{l}^{(n)}}{m+\frac{in}{R}},...  \big)\,.
\end{align}
Furthermore, the new neutrino mass matrix in the newly introduced basis will be
\begin{align}
\label{eq:uniTrans}
\mathcal{M}^R = (U^\nu)^T \mathcal{M}\,,
\end{align} 
where the first three rows vanish and give rise to massless chiral
fields and the remaining rectangular matrix refers to the massive
Dirac fields. In addition, since the chiral fields $\nu_l$ are
predominantly left-handed, in the case of $m_{l}^{(n)}\ll m$ according to \cite{Ioannisian:1999cw} we have
\begin{align}
\label{eq:neutLeft}
\nu_{iL}= (1_{3\times 3} + \Xi^* \Xi^T)^{-1/2} \big(\nu_l + \Xi^* \Psi_{+}^{'R}\big)
\end{align} 
such that $\nu_{l(L)}^{T}=(\nu_{e(L)}, \nu_{\mu(L)}, \nu_{\tau(L)})$
and $(\Psi_{+}^{'R})^T=(\eta_{0}^{R}, \eta_{1}^{R},\eta_{-1}^{R},...,\eta_{n}^{R},\eta_{-n}^{R},... )$. A good discussion for
the scenario $m\rightarrow 0$ can be found in \cite{Hamed}, where a
level crossing effect occurs that makes $\eta_0$ massless, but a
linear combination of the chiral fields $\nu_l$ obtains a tiny Dirac
mass of order $m_{l}^{(0)}$ while the other two linear combinations orthogonal to the first are still massless. 
Let us now focus on the diagonalization of the rectangular part of the matrix $\mathcal{M}$, denoted by $\mathcal{M}_\chi$. 
This matrix is spanned by the
fields $\Psi_{+}^{'R}$ and $\Psi_-$, and diagonalized using bi
unitary transformation $\mathcal{V}_{+}^{T} \mathcal{M}_\chi \mathcal{V}_-= \widehat{ \mathcal{M}_\chi }$, where $\mathcal{V}_+$
and $\mathcal{V}_-$ are unitary matrices.

After this process, the form of the diagonal matrix is given by \cite{Ioannisian:1999cw}
\begin{align}
\label{eq: diagMass}
 \widehat{ \mathcal{M}_\chi}\approx\begin{pmatrix}
m& 0& 0& \cdots& 0& 0&  \cdots\\
0& \sqrt{m^2 +\dfrac{1}{R^2}}& 0& \cdots& 0& 0& ... \\
0& 0& \sqrt{m^2 +\dfrac{1}{R^2}}& \cdots& 0& 0&  \cdots \\
\vdots& \vdots& \vdots& \ddots& \vdots& \vdots&  \cdots \\
0& 0& 0& \cdots& \sqrt{m^2 +\dfrac{n^2}{R^2}}& 0&  \cdots \\
0& 0& 0& \cdots& 0& \sqrt{m^2 +\dfrac{n^2}{R^2}}& \cdots \\
\vdots& \vdots& \vdots& \vdots& \vdots& \vdots&  \ddots
\end{pmatrix}\,,
\end{align}
up to the leading order of $m_{l}^{(n)}/m$. We call $\chi^{(n)}$ the
KK mass eigenfields corresponding to the diagonal entries of $\widehat{ \mathcal{M}_\chi}$. As noted in \cite{Ioannisian:1999cw,Dvali}, both  matrices $\mathcal{V}_+$ and $\mathcal{V}_-$ associated to this
approximation are close to the identity matrix up to phase factors.
One can directly observe from $\widehat{ \mathcal{M}_\chi}$ that all
KK mass eigenstates $\chi^{(n)}$ occur in degenerate pairs, i.e. $m_{(-n)}=m_{(n)}$, except for $\chi^{(0)}$ with mass $m_{(0)}\approx m$. Therefore, the KK state $\chi^{(0)}$ is the second lightest among neutrino states as $\nu_l$ is still massless~\cite{Ioannisian:1999cw}.

The following expressions are the interaction Lagrangians involving the neutrino states $\nu_l$ and $\chi^{(n)}$, the charged leptons $l$, 
as well as the weak vector bosons $Z$ and $W^{\pm}$, and their respective Goldstone boson candidates $G^0$ and $G^{\pm}$~\cite{Schter}:
\begin{align}
\label{eq: I1}
\mathcal{L}_{Z}^{int} = &-\dfrac{g_w}{2c_w} Z^\mu\Big[ \sum_{l, l'=e, \mu, \tau} C_{\nu_l, \nu_{l'}} \bar{\nu}_l \gamma_\mu P_L \nu_{l'} + \Big( \sum_{l=e, \mu, \tau} \sum_{n=-\infty}^{\infty} C_{\nu_l, n} \bar{\nu}_l \gamma_\mu P_L \chi^{(n)}    + h.c. \Big) \nonumber \\
& +\sum_{n,k=-\infty}^{\infty} C_{n,k} \bar{\chi}^{(n)} \gamma_\mu P_L \chi^{(k)}     
\Big]\,, \nonumber \\
\mathcal{L}_{W^{\pm}}^{int}&=-\dfrac{g_w}{\sqrt{2}} W^{-\mu} \sum_{l=e, \mu, \tau}
\Big( B_{l\nu_l}\bar{l}\gamma_\mu P_L \nu_l + \sum_{n=-\infty}^{\infty} 
 B_{l, n}\bar{l}\gamma_\mu P_L \chi^{(n)} + h.c. \Big)\,, \nonumber \\
\mathcal{L}_{G^0}^{int} &= \dfrac{ig_w}{2M_W} G^0 \Big[\sum_{l=e, \mu, \tau} \Big(
\sum_{n=-\infty}^{\infty} C_{\nu_l,n} m_{(n)}\bar{\nu}_l P_R \chi^{(n)}+ h.c.      \Big)  \nonumber \\
& -\sum_{n,k=-\infty}^{\infty} C_{n, k}  \bar{\chi}^{(n)} \big( m_{(n)}P_L - 
m_{(k)}P_R \big)\chi^{(k)}  \Big]\,,  \nonumber \\
\mathcal{L}_{G^\pm}^{int} &= -\dfrac{g_w}{\sqrt{2} M_W} G^-  \sum_{l=e, \mu, \tau}
\Big[B_{l\nu_l}m_l\bar{l} P_L \nu_l + \sum_{n=-\infty}^{\infty}  B_{l, n}\bar{l}
\big(m_l P_L - m_{(n)}P_R\big) \chi^{(n)} + h.c.\Big]\,,
\end{align}  
where $P_{R, L}=\dfrac{1\pm\gamma_5}{2}$ are the chirality
projection operators and $g_w$ is the weak coupling constant. The $m_{(n)}$ and $m_l$ denote the mass of the $\rm{n^{th}}$ KK state
and charged leptons. The expressions for the matrix elements of $B$
are given by \cite{Ioannisian:1999cw}
\begin{align}
\label{eq: VU}
 B_{l,n}=\sum_{i=e, \mu, \tau} V_{li}^{l} U_{i,n}^{\nu}\,,\qquad B_{l\nu_k}= \sum_{i=e, \mu, \tau} V_{li}^{l} U_{ik}^{\nu}  \quad \textrm{for each}\:\: k=e, \mu, \tau\,,  
\end{align}
while the matrix elements of $C$ are described by
\begin{align}
\label{eq: BB}
C_{\nu_l,n}= \sum_{k=e, \mu, \tau} B_{k\nu_l} B_{k, n}^{*}, \quad 
C_{n,m}= \sum_{k=e, \mu, \tau} B_{k,n} B_{k, m}^{*}\,, 
\quad C_{\nu_l\nu_{l'}}= \sum_{k=e, \mu, \tau} B_{k\nu_l} B_{k \nu_{l'}}^{*}\,.
\end{align}
Here, $V^l$ is the unitary matrix that diagonalizes the mass matrix
of the charged leptons. At the tree level and beyond, the
interaction Lagrangians in Eq.~\eqref{eq: I1} generate new physics
effects. For this reason, we can study the contributions of the KK
neutrinos to the electroweak observables taking place at the tree
level and beyond. Consequently, the bound on the mixing parameters and the
fundamental scale $M_F$ can also be estimated. In this case, the
mixing parameters coming from the mixing angles with all of the KK neutrinos are given by \cite{LangLon, Ioannisian:1999cw} 
\begin{align}
\label{eq: mixPar}
(s_{L}^{\nu_l})^2 =\sum_{n=-\infty}^{\infty} |B_{l,n}|^2=[V^l ( 1_{3\times 3} + \Xi^* \Xi^T )^{-1/2} (\Xi^* \Xi^T )( 1_{3\times 3} + \Xi^* \Xi^T )^{-1/2}V^{l\dagger} ]_{l\:l} .
\end{align}
Without loss of generality, we assume that the charged lepton mass
matrix has only non-negative and diagonal entries such that $V^l$ is 
given by the identity matrix. As a consequence, the mixing parameter is
approximated to $(s_{L}^{\nu_l})^2 \approx [\Xi^* \Xi^T]_{l\:l}$ up
to the leading order of $(h_l v)/ M_F$. 
In addition, the discrete summation over the KK modes can be
replaced by the integration over the energy scale $E$~\cite{Ioannisian:1999cw}
\begin{align}
\label{eq: replacement}
\sum_{n=-\infty}^{\infty} \longrightarrow S_\delta R^\delta \int_{0}^{M_F} E^{\delta-1}dE,
\end{align}
where $M_F$ is the ultraviolet (UV) cutoff, $R$ is the radius of the circle in the compactification,  and  $S_\delta=\dfrac{2\pi^{\frac{\delta}{2}}}{\Gamma(\frac{\delta}{2})}$ denotes the surface area of the unit sphere in $\delta$ dimensions. As a result, the mixings can be simplified as  
\begin{equation}
\label{eq: formMix}
(s_{L}^{\nu_l})^2\approx [\Xi^* \Xi^T]_{l\:l}= \begin{cases} 
      \dfrac{\pi h_{l}^{2}v^2 }{2 M_{F}^{2} } \ln \Big( \dfrac{m^2 + M_{F}^{2}}{m^2}\Big)
& \textrm{for}\:\:\: \delta=2\\
\dfrac{S_\delta}{\delta-2} \Big(\dfrac{h_{l}^{2}v^2}{2M_{F}^{2}} \Big) \Big( 1 + \mathcal{O}\left(\dfrac{m^2}{M_{F}^{2}}\right) \Big)  & \textrm{for}\:\:\: \delta>2. 
   \end{cases}
\end{equation}     


\section{$b\to s l l$ transition and constraints}
 \label{sec:numerical}
\begin{figure}
	\centering
	\includegraphics[width=0.9\textwidth]{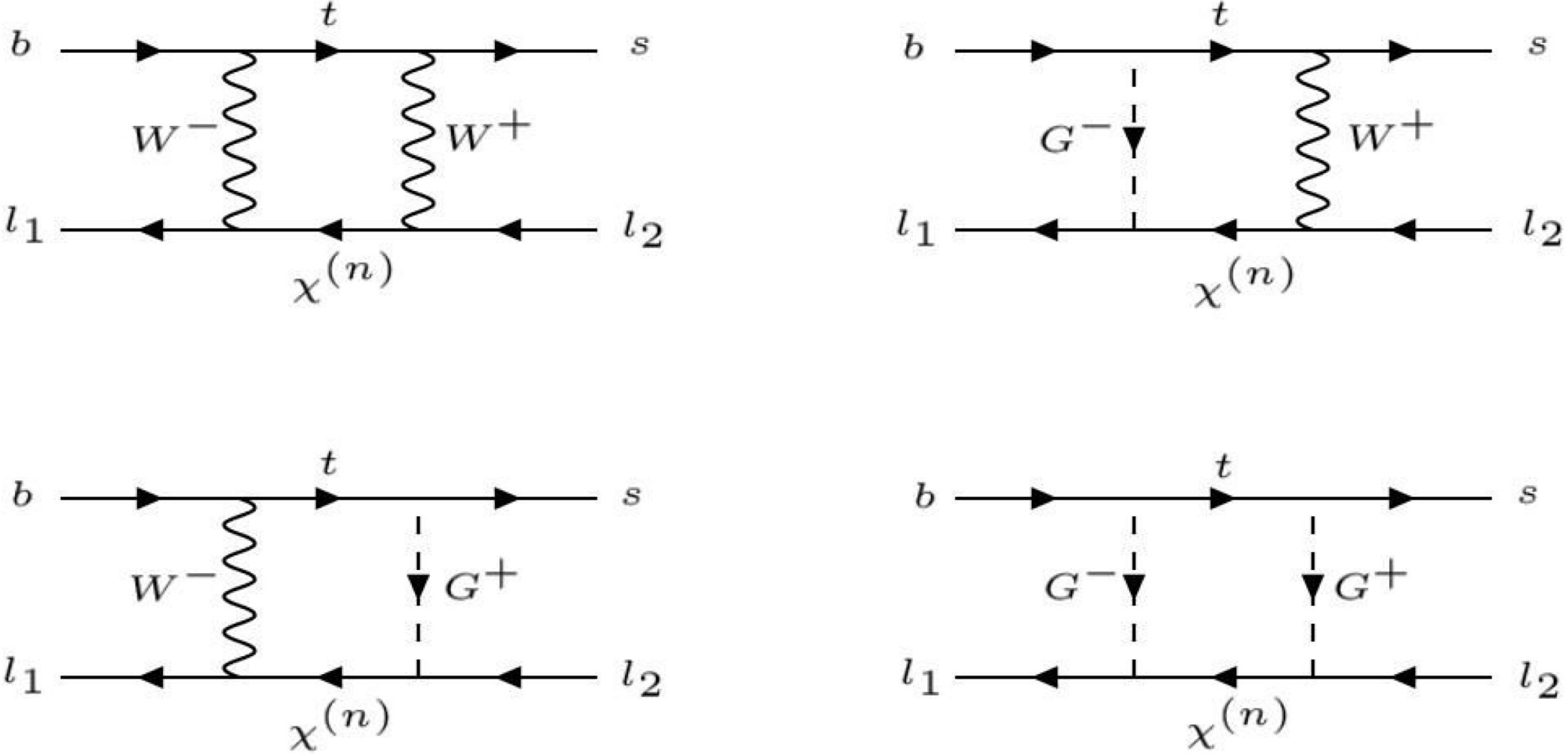}
	\caption{The relevant Feynman diagrams for $b \rightarrow s l_{1} \bar{l}_{2}$ transition.}
	\label{fig:bsl1l2}
\end{figure}

The corresponding Feynman diagram of $b \rightarrow s l_{1} \bar{l}_{2}$ transition in this model is given by the box diagram in Fig.\ref{fig:bsl1l2}. 
This can be written in general amplitude expression as~\cite{Ioannisian:1999cw}
\begin{align}
\label{eq:bsllAmplitude}
\mathcal{T}_{\text{box}}(b\rightarrow sl_{1}\bar{l}_{2}) = -\frac{i\alpha^{2}_{w}}{16M^{2}_{W}} F^{bsl_{1}l_{2}}_{\text{box}} \bar{u}_{s} \gamma_{\mu} (1-\gamma_{5}) u_{b} \,\, \bar{u}_{l_{1}} \gamma^{\mu} (1-\gamma_{5}) v_{l_{2}}\, ,
\end{align} 
where $\alpha=\frac{g_w^2}{4\pi}$ with $g_w$ being the weak coupling strength and $M_W$ is the mass of $W$-boson. The composite form factor $F^{bsl_{1}l_{2}}_{\text{box}}$ appears in this equation is given by
\begin{align}
\label{eq:Fboxbsll}
F^{bsl_{1}l_{2}}_{\text{box}}=V^{*}_{ts} V_{tb}\,\sum^{n=\infty}_{n=-\infty}  B^{*}_{l_{2},n}\, B_{l_{1},n}\, \left[F_{\text{box}}(\lambda_{t},\lambda_{n})-F_{\text{box}}(0,\lambda_{t})-F_{\text{box}}(0,\lambda_{n})+F_{\text{box}}(0,0)\right]\,,
\end{align}
where $V_{ts}$ and $V_{tb}$ are elements of CKM matrix,  $\lambda_{t}$ and $\lambda_{n}$ stand for $m^{2}_{t}/M^{2}_{W}$ and $m^{2}_{(n)}/M^{2}_{W}$, respectively.
The analytic form of the loop function $F_{\text{box}}(x,y)$ reads~\cite{Ioannisian:1999cw,Ilakovac:1994kj}
\begin{align}
\label{eq:Fbox}
F_{\text{box}}(x,y)&=\frac{1}{x-y}\left[ \left(1 + \frac{xy}{4} \right)\left(\frac{1}{1-x} +\frac{x^{2}\ln x}{(1-x)^{2}}-\frac{1}{1-y}-\frac{y^{2} \ln y}{(1-y)^{2}} \right)\right]\nonumber \\ 
&- \frac{2xy}{x-y} \left[\frac{1}{1-x} +\frac{x\ln x}{(1-x)^{2}}-\frac{1}{1-y}-\frac{y\ln y}{(1-y)^{2}} \right]\,.
\end{align}
The dominant contribution of the loop function $F_{\text{box}}(\lambda_{t},\lambda_{n})$ comes from KK neutrinos heavier than W boson, $m^{2}_{(n)} >> M^{2}_{W}$. Taking this fact into account, the composite form factor $F^{bsl_{1}l_{2}}_{\text{box}}$ becomes
\begin{align}
\label{eq:FboxbsllAppr}
F^{bsl_{1}l_{2}}_{\text{box}} \approx V^{*}_{ts} V_{tb}\,\sum^{n=\infty}_{n=-\infty}  B^{*}_{l_{2},n}\, B_{l_{1},n}\, \left[K(\lambda_{t}) + \frac{\lambda_{t}}{4} \ln \lambda_{n} \right]\,,
\end{align}
where we have defined $n$ independent $K(\lambda_{t})$ function as
\begin{align}
\label{eq:K}
K(\lambda_{t}) = 1 + \frac{1}{1-\lambda_{t}}\,\left(\frac{7}{4}\lambda_{t}-1 \right) + \frac{\lambda_{t}}{(1-\lambda_{t})^{2}}\,\left[\lambda_{t}\left(2-\frac{\lambda_{t}}{4} \right)-1 \right] \ln \lambda_{t}\,.
\end{align}

To simplify the discrete summation over the $n$ index, one employs
the continuation limit, replacing the summation by the integral
as in Eq.\eqref{eq: replacement}. The final form of  $F^{bsl_{1}l_{2}}_{\text{box}}$ reads
\begin{align}
\label{eq:FboxsllFinal}
F^{bsl_{1}l_{2}}_{\text{box}}=V^{*}_{ts} V_{tb}\, s^{\nu_{l_{1}}}_{L}\, s^{\nu_{l_{2}}}_{L}\,\left[K(\lambda_{t}) + \frac{\lambda_{t}}{4}\, e_{\delta}(m,M_{F}) \right]\,,
\end{align}
where the $e_{\delta}$ function is separated into two parts according to different $\delta$ as
\begin{equation}
\label{eq:edelta2}
e_{\delta}(m,M_{F}) =I_\delta \begin{cases} 
      \dfrac{M^{2}_{F}}{\ln \left( \frac{M^{2}_{F}+m^{2}}{m^{2}} \right)}
& \textrm{for}\:\:\: \delta=2\\
 \dfrac{M^{2}_{F}}{2}\, (\delta-2)  & \textrm{for}\:\:\: \delta>2\,, 
   \end{cases}
\end{equation}    
%
where the discrete summation over the KK modes is encoded within the
integral function
\begin{align}
\label{eq:integral}
I_{\delta} = \int^{1}_{\frac{m^{2}}{M^{2}_{F}}} dw \frac{w^{\frac{\delta}{2}-1}}{M^{2}_{F}\,w + m^{2}}\, \ln\left( \frac{M^{2}_{F}\,w + m^{2}}{M^{2}_{W}} \right)\,.
\end{align}
Here, $w$ is the dimensionless integration variable obtained after changing the integration boundary as prescribed in~\cite{Ioannisian:1999cw}.
Note that for $\delta=2$ when $m=0$, one should replace the
logarithm expression in the denominator of Eq.~\eqref{eq:edelta2} by $\ln \left(M^{2}_{P}/M^{2}_{F}\right)$~\cite{Ioannisian:1999cw}.
In this paper, we take $\delta = 2, 3, 4, 5, \text{and}\,6$ to see
the implications of  the lower bound for  the fundamental scale $M_{F}$ based on the available experimental data. Since $m<< M_{F}$ the integral fucntion $I_{\delta}$ can be approximated as
\begin{align}
\label{eq:I2}
I_{2} \approx \frac{1}{2M^{2}_{F}} \left[ \ln \left(\frac{M^{2}_{F}+m^{2}}{m^{2}} \right) \right]^{2},
\end{align}
\begin{align}
\label{eq:I3I4}
I_{3} \approx \frac{2}{M^{2}_{F}}\left[ -2 + \ln\left( \frac{M^{2}_{F}}{M^{2}_{W}} \right) \right],\,\,\,\, I_{4} \approx \frac{1}{M^{2}_{F}}\left[ -1 + \ln\left( \frac{M^{2}_{F}}{M^{2}_{W}} \right) \right],
\end{align}
\begin{align}
\label{eq:I5I6}
I_{5} \approx \frac{2}{9 M^{2}_{F}}\left[ -2 + 3\ln\left( \frac{M^{2}_{F}}{M^{2}_{W}} \right) \right],\,\,\,\, I_{6} \approx \frac{1}{4M^{2}_{F}}\left[ -1 +2 \ln\left( \frac{M^{2}_{F}}{M^{2}_{W}} \right) \right].
\end{align}
One can see that the dependence of $m$ only shows up in the case of $\delta=2$. This is consistent with the Ref.~\cite{Ioannisian:1999cw}. Also notice that
the amplitude of $b \rightarrow s l_{1} \bar{l}_{2}$ transition
receives the logarithmic enhancement of $M^{2}_{F}$ which violates
the decoupling theorem~\cite{Appelquist:1974tg}.

In terms of effective Hamiltonian, the transition amplitude of $b \rightarrow s l \bar{l}$ ($l_{1} = l_{2} = l$) relevant for $R_{K}$ and $R_{K^{*}}$  can be written as~\cite{Fajfer:2012vx} 
\begin{align}
\label{eq:Heff}
\mathcal{H}_{\text{eff}} = -\frac{4G_{F}}{\sqrt{2}} \, \frac{\alpha}{4\pi}\,V^{*}_{ts} V_{tb} \sum_{i=9,10} C_{i}\mathcal{O}_{i} + \text{h}.\text{c}. \,,
\end{align}
where $G_F$ and $\alpha$ are the Fermi coupling constant and fine-structure constant, respectively. The corresponding operators are
\begin{align}
\label{eq:operator}
\mathcal{O}_{9} = (\bar{s}\gamma^{\mu} P_{L} b) (\bar{l}\gamma_{\mu}l) \,, \, \, \mathcal{O}_{10} = (\bar{s}\gamma^{\mu} P_{L} b) (\bar{l}\gamma_{\mu} \gamma_{5} l)\,.
\end{align}
The new physics contributions originated from the KK neutrinos running in the loops are encoded in the Wilson coefficients $C^{bsll}_{9}$ and $C^{bsll}_{10}$ as
\begin{align}
\label{eq:C9C10}
C^{bsll}_{9} = -C^{bsll}_{10} = \frac{(s^{\nu_{l}}_{L})^{2}}{16\, s^{2}_{w}} \left[4K(\lambda_{t}) + \lambda_{t}\, e_{\delta}(m, M_{F}) \right]\,,
\end{align}
where $s_{w}$ denotes the Weinberg angle. The constraints on $C^{bsll}_{9}$ and $C^{bsll}_{10}$ from $B$ anomalies are given by
the experimental values of several observables. Based on the results
of global fitting in all rare $B$ decays, the largest pull away the SM values indicates
that new physics hides in muonic sector. The preferred values of $C^{bsll}_{9}$ and $C^{bsll}_{10}$ are $C^{bs\mu\mu}_{9} = - C^{bs\mu \mu}_{10} = -0.41^{+0.07}_{-0.07}$~\cite{Altmannshofer:2021qrr}. Evidently, Eq.~\eqref{eq:C9C10} shows that the KK neutrino theory provides a natural explanation of these anomalies. Following the experimental
data of $B$ anomalies, we assume that the new physics resides in the
muon sector. 

\begin{figure}[t]
	\centering
\includegraphics[width=0.7\textwidth]{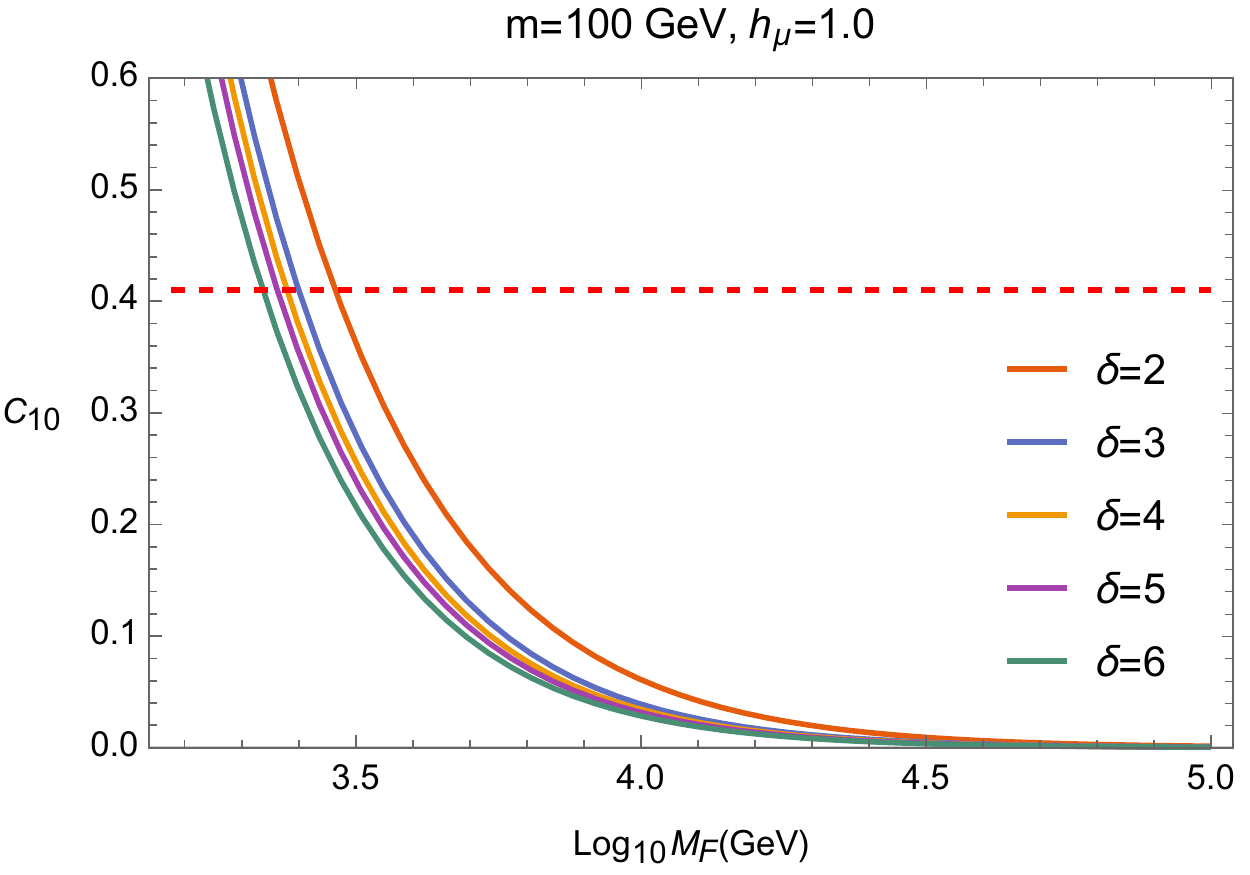}
	\caption{$C^{bs\mu\mu}_{10}=-C^{bs\mu\mu}_{9}$ as function of $M_{F}$ for $m=100$ GeV and $h_{\mu} = 1.0$ for different delta. The horizontal red dashed line at 0.41 is the central value of $C^{bs\mu\mu}_{10}$ from the global fits.}
	\label{fig:C10vsMFm100h1}
\end{figure}
\begin{figure}[t]
	\centering
	\includegraphics[width=0.45\textwidth]{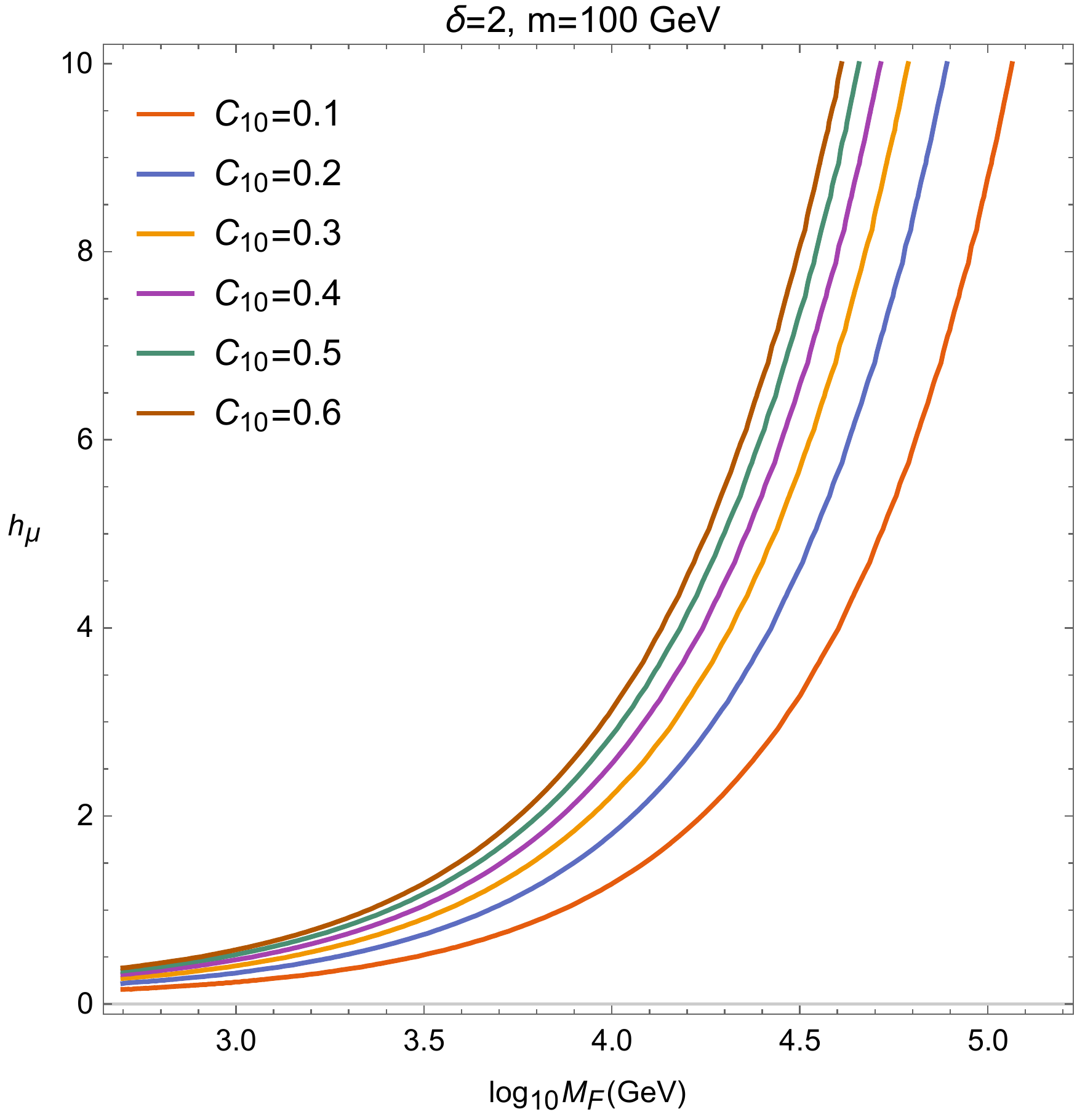}
	\includegraphics[width=0.45\textwidth]{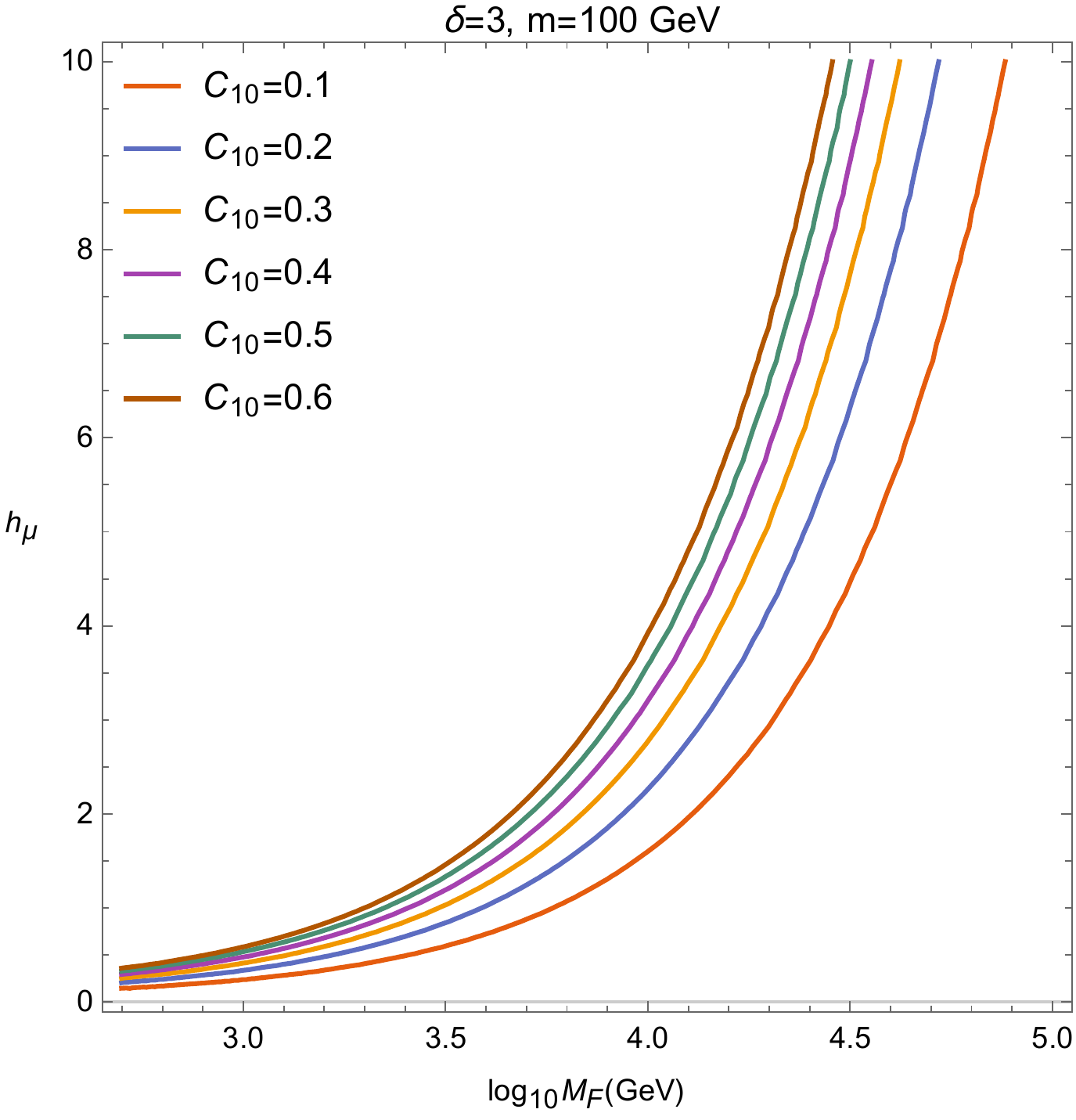}
	\caption{The contour of $h_{\mu}$ vs $M_{F}$ for $m = 100$ GeV. The left panel is for $\delta = 2$ and the right panel is for $\delta=3$. The contour plots  for higher $\delta$ are similar to the case of $\delta=3$.}
	\label{fig:CPhvsMFdelta2345}
\end{figure}

The relation between $C^{bs\mu\mu}_{10}=-C^{bs\mu\mu}_{9}$ and the fundamental scale $M_{F}$ for $m = 100$ GeV and $h_{\mu} = 1.0$ for $\delta = 2 \sim 6$ is shown in Fig.~\ref{fig:C10vsMFm100h1}.
The central value of $C^{bs\mu\mu}_{10}$ from the best fit ($C^{bs\mu\mu}_{10}$= 0.41) is denoted by the horizontal red dashed line. 
The interceptions between the horizontal red dashed line with the solid curves ($M_F= 2.17 \sim 2.91~\rm{TeV}$
) 
indicate what the fundamental scale should be in order to provide the explanation for the $R_K$ and $R_{K^*}$ anomalies  for different numbers of extra dimension $\delta$.
One can see that as $\delta$ gets smaller the fundamental scale $M_{F}$ becomes higher.
\begin{figure}[t]
	\centering
	\includegraphics[width=0.45\textwidth]{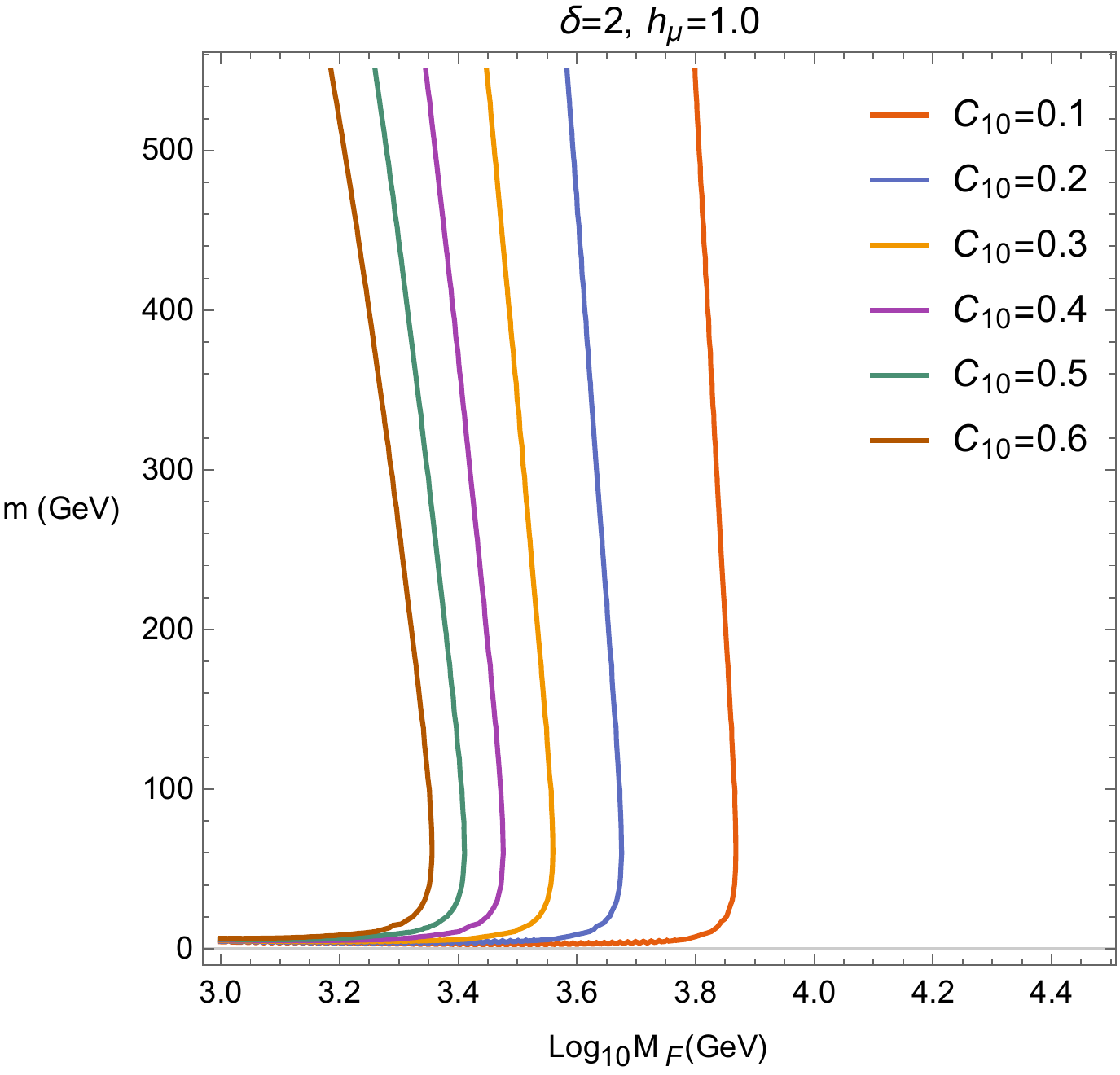}
	\includegraphics[width=0.45\textwidth]{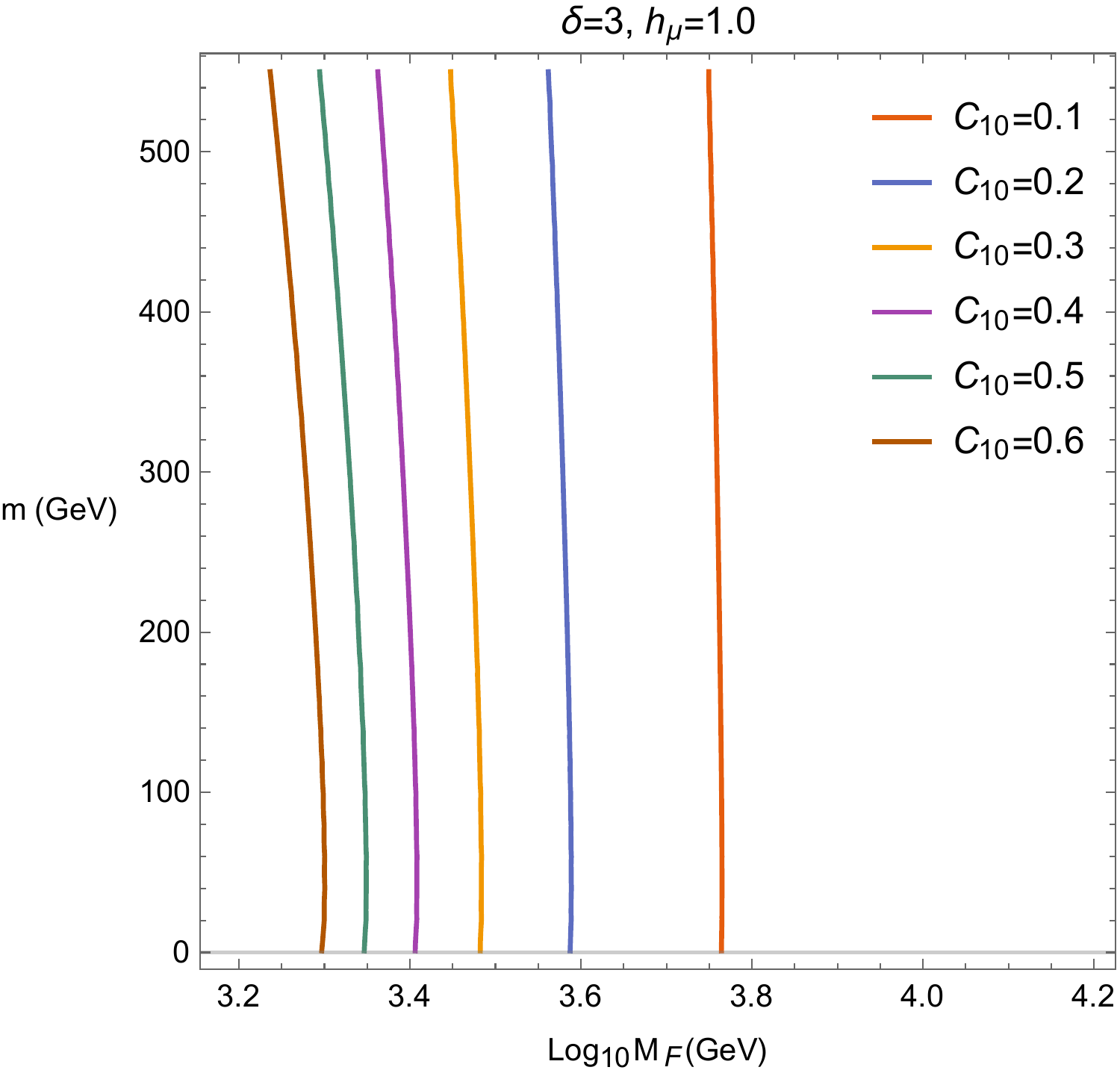}
	\caption{The contour of $m~\text{vs}~M_{F}$ for a fixed Yukawa coupling $h_\mu=1$. The left panel is for $\delta = 2$ and the right panel is for $\delta=3$. The contour plots  for higher $\delta$ are similar to the case of $\delta =3$.}
	\label{fig:CPmvsMFh1}
\end{figure}  
The dependence of $C^{bs\mu \mu}_{10}$ on $m$ for different $\delta$
are shown in Fig.\ref{fig:CPmvsMFh1}. It is clear
that  the $C^{bs\mu \mu}_{10}$ is slightly 
sensitive to the value of $m$ only when $\delta$ equals 2. This can be seen directly from Eq.\eqref{eq:I2} that for $\delta = 2$ the  $C^{bs\mu \mu}_{10}$ is
proportional to $\ln\left(\frac{M^{2}_{F} + m^{2}}{m^{2}} \right)$
while it is independent of $m$ for $\delta > 2$, c.f. Eq.~\eqref{eq:I3I4} and \eqref{eq:I5I6}. 
%
Finally, the relation between the fundamental scale $M_{F}$ and Yukawa coupling $h_{\mu}$ is given in Fig.~\ref{fig:CPhvsMFdelta2345} for $m = 100$ GeV. As expected, the higher Yukawa coupling corresponds to the higher limit on $M_{F}$. 

\begin{figure}
	\centering
	\includegraphics[width=0.8\textwidth]{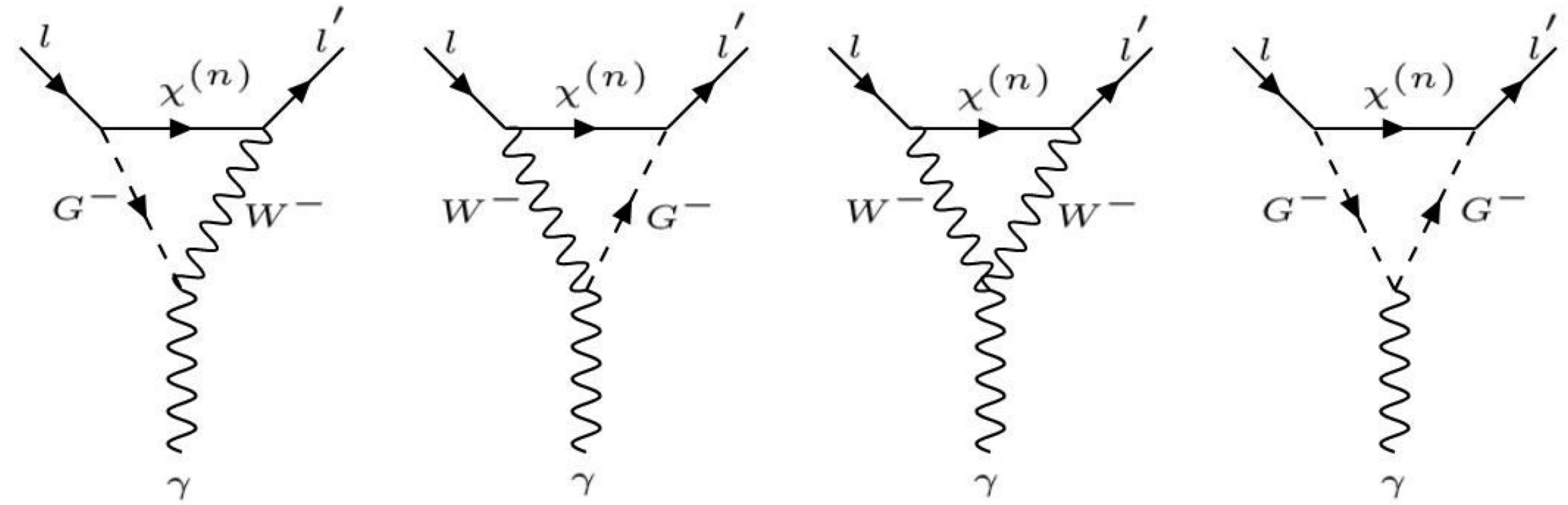}
	\caption{The relevant Feynman diagrams for $l \rightarrow l^{'} \gamma$ transition.}
	\label{fig:mueg}
\end{figure}

On the other hand, since the KK neutrinos only interact with
leptonic sector, one has to confront the experimental constraints
from this sector.  The most stringent limit is given by the lepton
flavor
violation (LFV) decay of muon to electron and photon. The Feynman
diagrams relevant for this process are shown in Fig.\ref{fig:mueg}.
The
general expression for the transition amplitude of $l(p_{l}) \rightarrow l^{'}(p_{l^{'}}) \gamma(q)$ is given by~\cite{Ioannisian:1999cw}
\begin{align}
\label{eq:Tmueg}
\mathcal{T}(l \rightarrow l^{'} \gamma) = \frac{i\,e\,\alpha_{w}}{16\pi \,M^{2}_{W}} \, G^{ll^{'}}_{\gamma} \epsilon^{\mu}_{\gamma}\, \bar{u}_{l^{'}} i\sigma_{\mu \nu} q^{\nu} \left[ m_{l^{'}} (1+\gamma_{5}) + m_{l} (1-\gamma_{5})\right] u_{l}\,. 
\end{align}
The composite form factor $G^{ll^{'}}_{\gamma}$ is written in terms of summation over the KK modes as~\cite{Ioannisian:1999cw}
\begin{align}
\label{eq:Gllp}
G^{ll^{'}}_{\gamma} = \sum^{n=\infty}_{n=-\infty} B^{*}_{l,n}\, B_{l^{'},n} \,G_{\gamma}(\lambda_{n})\,,
\end{align}
where the loop function $G_{\gamma}(\lambda_{n})$ is given by
\begin{align}
\label{eq:GllLoop}
G_{\gamma}(x) = - \frac{2x^{3}+5x^{2}-x}{4(1-x)^{3}} - \frac{3x^{3} \ln x}{2(1-x)^{4}}\,.
\end{align}
In the limit of $\lambda_{n} >> 1$, the composite form factor $G^{ll^{'}}_{\gamma}$ becomes
\begin{align}
\label{eq:GllpF}
G^{ll^{'}}_{\gamma} \approx \frac{1}{2} s^{\nu_{l}}_{L}\, s^{\nu_{l^{'}}}_{L}\,.
\end{align}
The branching ratio of $l \rightarrow l^{'}\gamma$ at one-loop is finally given by
\begin{align}
\label{eq:Bllpg}
Br(l\rightarrow l^{'} \gamma) = \frac{\alpha^{3}_{w}\, s^{2}_{w}}{256\pi^{2}} \, \frac{m^{4}_{l}}{M^{4}_{W}}\, \frac{m_{l}}{\Gamma_{l}} \, |G^{ll^{'}}_{\gamma}|^{2} \approx  \frac{\alpha^{3}_{w}\, s^{2}_{w}}{1024\pi^{2}} \, \frac{m^{4}_{l}}{M^{4}_{W}}\, \frac{m_{l}}{\Gamma_{l}} \, (s^{\nu_{l}}_{L})^{2}\, (s^{\nu_{l^{'}}}_{L})^{2}\,.
\end{align}
For $\mu \rightarrow e \gamma$ transition, the corresponding branching ratio is
\begin{align}
\label{eq:Bmueg}
Br(\mu \rightarrow e \gamma) =  \frac{\alpha^{3}_{w}\, s^{2}_{w}}{1024\pi^{2}} \, \frac{m^{4}_{\mu}}{M^{4}_{W}}\, \frac{m_{\mu}}{\Gamma_{\mu}} \, (s^{\nu_{\mu}}_{L})^{2}\, (s^{\nu_{e}}_{L})^{2}\,.
\end{align}
Using the current experimental value from MEG experiment, $Br_{\text{exp}}(\mu \rightarrow e \gamma) < 4.2 \times 10^{-13}$~\cite{MEG:2016leq} at 90$\%$ confidence level (CL), one gets the following limit on $s^{\nu_{e}}_{L} \, s^{\nu_{\mu}}_{L}$
\begin{align}
\label{eq:snumue}
s^{\nu_{e}}_{L} \, s^{\nu_{\mu}}_{L} < 2.11 \times 10^{-5}\,.
\end{align}
Since we assume the new physics comes from muon sector, we set the
mixing angle in electron sector $s^{\nu_{e}}_{L}$ to be very small.
If we take electron Yukawa coupling $h_{e} = 10^{-4}$, $h_{\mu}=1.0$, and $m=100$ GeV, the current bound~\cite{MEG:2016leq}  requires 
\begin{align}
\label{eq:LimMFmueg}
M_{F} > 1.58,\, 1.34,\, 1.19,\, 1.12,\, 1.05\,\,\, \text{TeV}\,,
\end{align}
for $\delta$ equals to 2, 3, 4, 5, and 6,
respectively.    
We see that the lower bound of the fundamental scale $M_{F}$
obtained from this LFV constraint is compatible with the limit from
B anomalies as can be seen from Fig.~\ref{fig:C10vsMFm100h1}.

\section{Conclusions and discussions}
\label{sec:con}
Anomalies in rare decays of B mesons point to possible violation of lepton universality in the weak interaction, which is certainly a signal of physics beyond the SM. Recent measurement of $R_K$ by LHCb collaboration strengthens the hint of such a violation in $b\to s l^+  l^-$. In this paper, we show that it is possible to explain these anomalies in the extra-dimensional framework where the original Planck scale can be lowered to a fundamental scale $M_F$. With a SM singlet Dirac neutrino propagating in the bulk, the contributions from its KK modes provide a good explanation for the anomalies through the mixings with active neutrinos. For simplicity, we only consider the cause of anomalies due to the significant mixing with muon neutrino. To fit the favorite Wilson coefficients $C_9^{bs\mu\mu}=-C_{10}^{bs\mu\mu}=-0.41$ in the global analysis, we show that the fundamental scale $M_F$ can be  as low as $2.9$ TeV, assuming two spatial dimensions besides original $(3+1)$-dimensional spacetime and the muon Yukawa coupling strength $h_\mu=1$. From a conservative point of view, this energy scale can also be interpreted as the lower limit for $M_F$, assuming KK neutrinos only partially fill the gap between data and the SM prediction. Furthermore, when the number of extra dimensions increases, the preferred  $M_F$ tends to be lower. The main constraint comes from the rare $\mu \to e \gamma$ decay. However, since the mixing with electron neutrino is also involved, that sets a lower bound about $1.6$ TeV for $M_F$ if electron and muon Yukawa couplings are of ${\cal O}(10^{-4})$ and ${\cal O}(1)$, respectively, and number of extra dimensions is two. 
 
We would like to comment on the contributions of KK neutrinos to the anomalous magnetic
moment of muon, so-called $g_\mu -2$ before closing this paper.  Combing with BNL E821 data, the recent experimental value obtained by E989 at Fermilab
reaches a 4.24\,$\sigma$ deviation from the SM prediction~\cite{Muong-2:2021ojo,Aoyama:2020ynm}
\begin{align}
\label{eq:muong2}
\Delta a^{\text{exp}}_{\mu} = a^{\text{exp}}_{\mu} - a^{\text{SM}}_{\mu} = (251 \pm 59) \times 10^{-11}\,.
\end{align}
The corresponding expression of $\Delta a_{\mu}$ in KK model can be extracted from $\mu \rightarrow e \gamma$ transition amplitude
\begin{align}
\label{eq:KKamu}
\Delta a_{\mu}  \approx \frac{\alpha}{8\pi\,s^{2}_{w}} \left(\frac{m_{\mu}}{m_{W}} \right)^{2} \, (s^{\nu_{\mu}}_{L})^{2}\,.
\end{align}
In order to fit the central
value of the measured $\Delta a^{\text{exp}}_{\mu}$ from the Fermilab in Eq.~\eqref{eq:muong2}, one  obtains the value of $M_{F}$ within the
range between 450 GeV  and 524 GeV for different $\delta$ considered here with  $h_{\mu} = 1.0$ and $m = 100$ GeV.
%
This is in a serious tension with the needed value of $M_F$ for B
anomalies shown in Fig.~\ref{fig:C10vsMFm100h1}. In addition, such small values of $M_{F}$ in hundred GeV regime are disfavored by collider search for signals of extra dimensions. 
Therefore, if $g_\mu -2$ anomaly is confirmed in the future, the simple extension of SM in extra-dimensional framework we consider in this paper must be improved. 

\section*{Acknowledgment}  
This work was supported in part by the Ministry of Science and Technology (MOST) of Taiwan under Grant No.MOST 109-2112-M-003-004-, 110-2112-M-003-003- and 110-2811-M-003-505-.


\end{document}